\documentclass[apjl]{emulateapj}
\def\ergs{\hbox{ergs~s$^{-1}$}}

\begin{document}

\title{Spectral Evolution of NGC 1313 X-2: Evidence Against The Cool Disk Model}

\author{Hua Feng\altaffilmark{1} and Philip Kaaret\altaffilmark{1}}

\altaffiltext{1}{Department of Physics and Astronomy, The University of Iowa, Van
Allen Hall, Iowa City, IA 52242; Email: hua-feng@uiowa.edu}

\shortauthors{Feng and Kaaret}
\shorttitle{Spectral Evolution of NGC 1313 X-2}

\begin{abstract}
The presence of a cool multicolor disk component with an inner disk temperature $kT = 0.1 \sim 0.3$~keV at a luminosity $L > 10^{40}$~\ergs\  has been interpreted as evidence that the ultraluminous X-ray source NGC~1313 X-2 harbors an intermediate-mass black hole (IMBH). The temperature of a disk component should vary with luminosity as $L\propto T^4$. However, upon investigating the spectral evolution with multiple XMM-Newton observations, we found that the cool disk component failed to follow this relation with a confidence level of 0.999964. Indeed, the luminosity decreases as the temperature increases, and the luminosities at high temperatures are more than an order of magnitude less than expected from the $L\propto T^4$ extrapolation of luminosities at low temperatures.  This places a strong constraint against the validity of modeling the X-ray spectra of NGC~1313 X-2 as emission from the accretion disk of an IMBH.  The decrease in luminosity with increasing temperature of the soft component follows the trend suggested by a model in which the soft emission arises from an outflow from a stellar-mass black hole with super-Eddington accretion viewed along the symmetry axis. Alternatively, the spectra can be adequately fitted by a $p$-free disk model with $kT\approx2$~keV and $p\approx0.5$.  The spectral evolution is consistent with the $L\propto T^4$ relation and appears to be a high luminosity extension of the $L-kT$ relation of Galactic black holes. This, again, would suggest that the emission is from a super-Eddington accreting stellar mass black hole.
\end{abstract}


\section{Introduction}

Ultraluminous X-ray sources (ULXs) are pointlike, nonnuclear X-ray sources with luminosities above the Eddington limit of a 20 $M_\sun$ black hole \citep[$3\times10^{39}$ ergs s$^{-1}$; for a review, see][]{fab06}.  Strong variability from many ULXs indicates they are compact objects.  If the X-rays are emitted isotropically below the Eddington limit, then the high luminosity would indicate that ULXs contain IMBHs \citep{col99,kaa01}. However, the emission could be mechanically or relativistically beamed \citep{kin01,kor02}, or exceed the Eddington limit \citep{beg02,ebi03}; in these cases, IMBHs are not required. 

The sum of a power-law and a multicolor disk \citep[MCD;][]{sha73} has been found to describe well the X-ray spectra of black hole binaries.  Fitting ULX spectra with the same model often shows a disk component, first reported in \citet{kaa03}, with a disk temperature around 0.1--0.4 keV which is much lower than that found in stellar-mass black holes.  The cool-disk high-luminosity regime suggests that ULXs harbor IMBHs with masses of $10^3$~$M_\odot$ \citep{kaa03,mil03,fen05}.

The radial temperature of an MCD follows a power-law form as $kT\propto R^{-p}$, where $p=0.75$. When taking into account advective cooling, the power-law dependence $p$ is predicted to decrease until 0.5 \citep{wat00}, which usually happens at very high accretion rates when the source enters the `slim disk' or optically thick advection-dominated accretion disk phase \citep{abr88}. The $p$-free model is similar to the MCD model but allows $p$ to vary \citep{min94}, and has been successfully applied in interpreting the very high state of black hole binaries such as XTE J1550$-$564 \citep{kub04}.  \citet{vie06} found that four ULXs that were previously reported to contain cool disks, could also be adequately fitted using the $p$-free model with disk temperatures of around 2~keV and $p\approx0.5$. They appear in the $L-kT$ diagram as an extension of stellar-mass black holes with super-Eddington radiation.

NGC~1313 X-2 is one of the ULXs with a spectrum that can be modeled by a cool MCD \citep{mil03}, a hot MCD \citep{sto06}, or a hot $p$-free disk \citep{miz07,vie06}.  The totally opposite physical consequences derived from the cool and hot disk models inspire us to investigate whether one can be ruled out. The source resides in an optical nebular supershell \citep{pak03}, whose kinetic energy is much larger than a typical supernova explosion.  The shell is either a hypernova remnant or a continually powered nebula possibly driven by the outflow from the ULX.  

In this Letter, we revisited the 12 archival {\it XMM-Newton} observations described in \citet{fen06}.  We present the spectral analysis in \S\ref{sec:spec} and discuss the nature of the ULX in \S\ref{sec:diss}.  We adopt a distance to the host galaxy of 4.13 Mpc \citep{men02}.

\section{Spectral Analysis}
\label{sec:spec}

\citet{fen06} analyzed 12 {\it XMM-Newton} observations of NGC~1313 and described the spectral state transitions of the two ULXs X-1 and X-2. This previous Letter identified a tight, positive correlation between the luminosity and the power-law photon index of X-1, while X-2 presented an opposite pattern of high/hard to low/soft spectral evolution consistent with previous {\it ASCA} observations \citep{zam04}. 

\begin{figure}[t]
\centering
\plotone{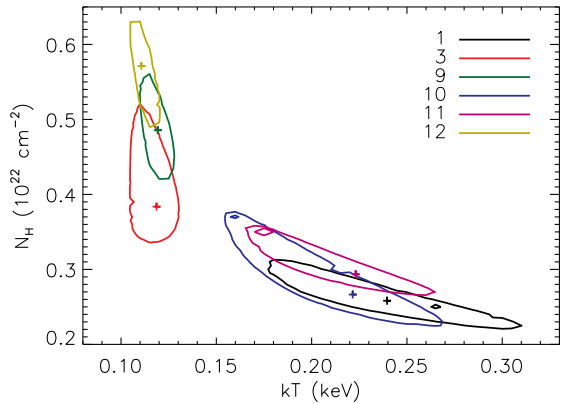} 
\caption{1 $\sigma$ confidence contours between the absorption column density $N_{\rm H}$ and the inner disk temperature $kT$ for the MCD model. Difference colors represent for different observations noted in \citet{fen06}. The plus signs indicate the best-fit values.  For each observation, the range of $kT$ in a contour is adopted as its error, used in Fig.~\ref{fig:diskbb}.
\label{fig:cont}}
\end{figure}

\begin{figure}[t]
\centering
\plotone{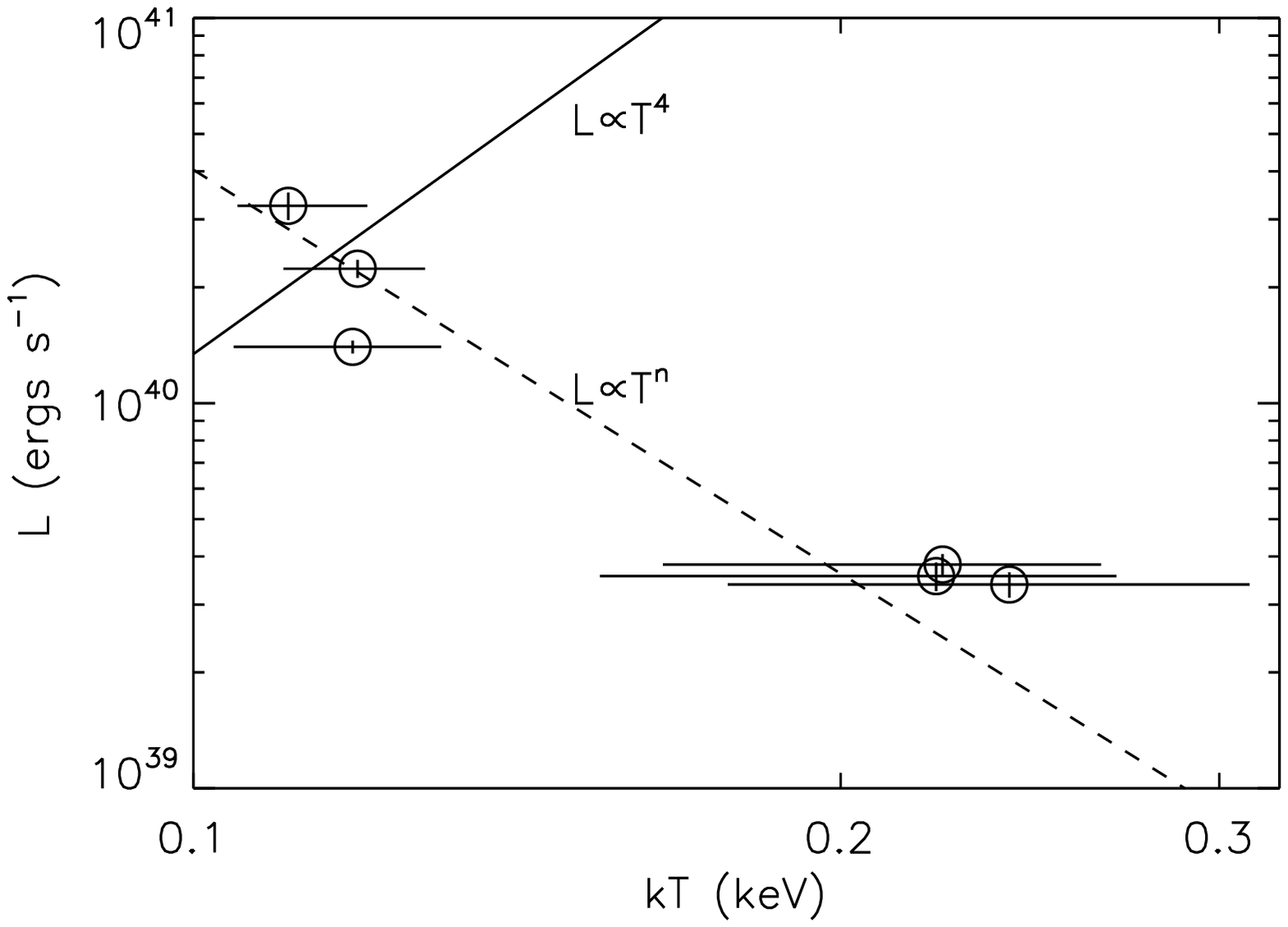} \\
\plotone{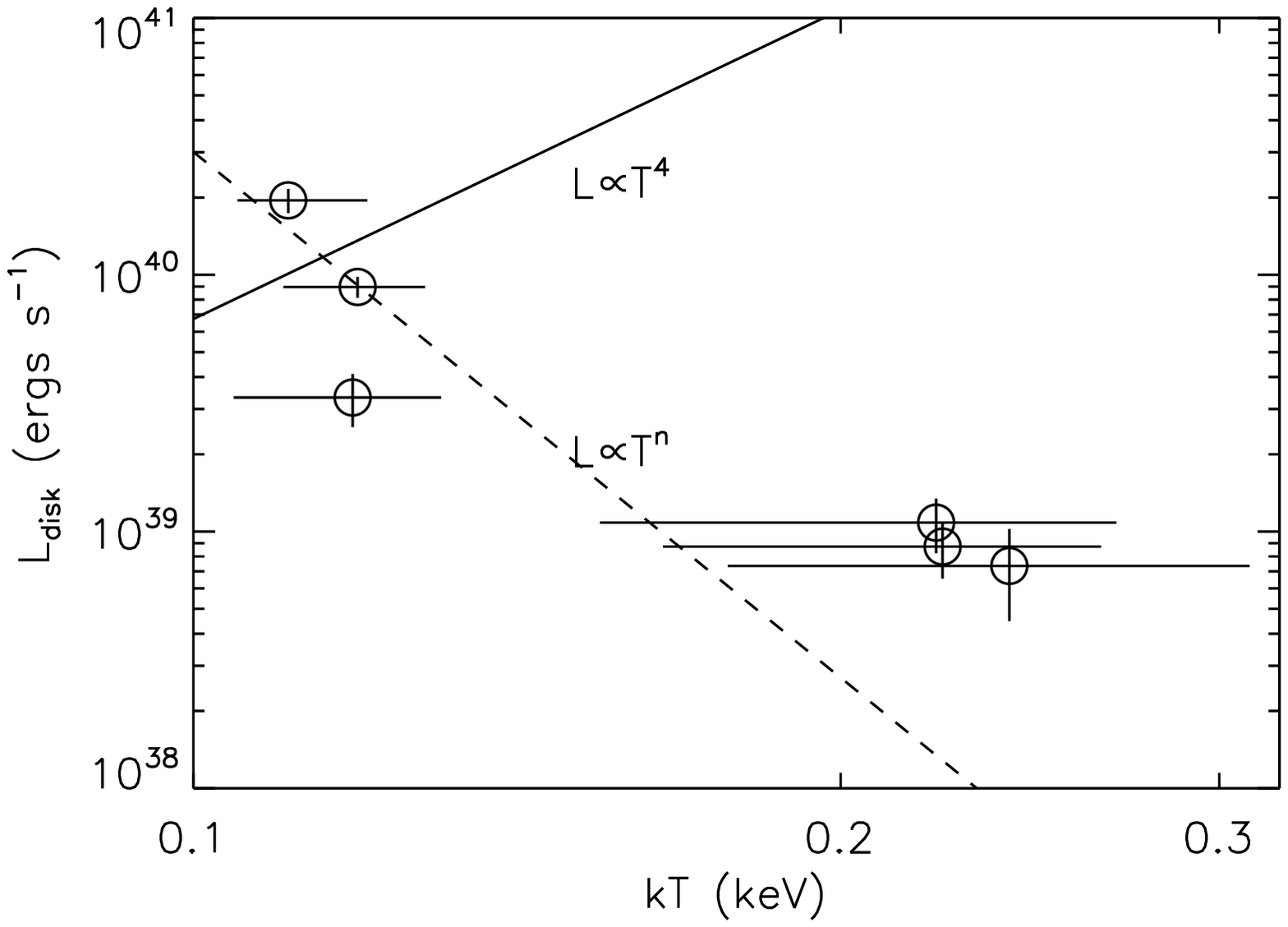} 
\caption{$L-kT$ diagram of NGC 1313 X-2 for the MCD model, with total 0.3--10 keV luminosity ({\it top}) and 0.3--10 keV disk luminosity ({\it bottom}). Solid lines indicate the best-fit $L \propto T^4$ relation, with $\chi^2/{\rm dof}=28/5$ for the $L-kT$ diagram and $\chi^2/{\rm dof}=41/5$ for the $L_{\rm disk}-kT$ diagram. Dashed lines are the best-fit $L \propto T^n$ relation, where $n=-3.7\pm0.7$ for the $L-kT$ diagram and $n=-8\pm2$ for the $L_{\rm disk}-kT$ diagram. 
\label{fig:diskbb}}
\end{figure}

\begin{figure}[t]
\centering
\plotone{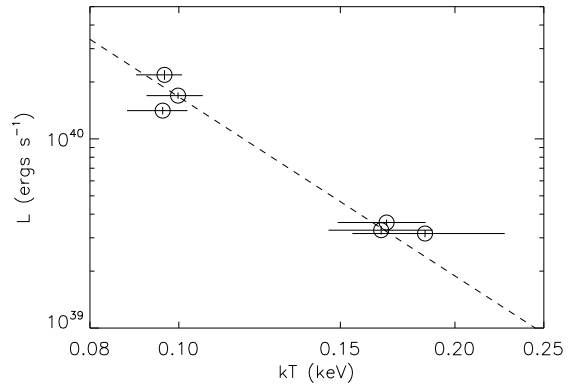} 
\caption{$L-kT$ diagram of NGC 1313 X-2 for the blackbody model. The dashed line indicates the best-fit $L \propto T^n$ relation, where $n=-3.1\pm0.5$ and $\chi^2/{\rm dof}=2.7/4$. 
\label{fig:bb}}
\end{figure}

In 6 of the 12 observations, NGC 1313 X-2 exhibits a significant soft excess which can be fitted by an MCD model with temperatures of 0.1--0.3 keV, contributing 20\%--60\% to the total 0.3--10 keV flux. All spectral parameters are listed in Table~1 of \citet{fen06}. Because the MCD component peaks where the interstellar absorption is of high importance and the observed MCD flux is often low relative to the power-law flux, we have carefully reevaluated the error on the temperature $kT$ to include the influence of the absorption column density $N_{\rm H}$, rather than using the default 1 $\sigma$ value derived from the spectral fits in \citet{fen06}. We created two-dimensional plots of the 1 $\sigma$ confidence contours allowing both $N_{\rm H}$ and $kT$ to vary (see Fig.~\ref{fig:cont}).  The contours are calculated on a grid of points with step sizes of 0.005~keV and $5 \times 10^{19}$~cm$^{-2}$.  To find the error on $kT$, we use the full range in $kT$ covered by the 1 $\sigma$ confidence range, i.e.\ from the leftmost to the rightmost point of the contour.  This error includes the influence of $N_{\rm H}$ and is larger than the conventional definition of the error derived from the {\tt error} command in XSPEC.  For these six observations, the 0.3--10 keV total luminosity versus temperature is plotted in Figure~\ref{fig:diskbb} (top), while the 0.3--10 keV disk luminosity versus temperature is plotted in the bottom panel.

We refitted all of the observations using a {\tt comptt} model in place of the power-law, and the MCD temperature changed by less than 0.02~keV in all cases.  Replacing the power-law model by a Comptonization model has little effect on the MCD component, because the source spectrum does not exhibit curvature or a break at high energies \citep{sto06}.  We also fitted the data with a model consisting of a simple blackbody and a power-law.  The results are shown in Figure~\ref{fig:bb}.  The error on the temperature is derived in the same way as for the MCD component, including the influence of the absorption on the error estimate.

To see if the data are consistent with a $L\propto T^4$ relation, we present a best-fit $L\propto T^4$ line on each plot. Since the uncertainty on the temperature is much larger than on the luminosity, we treated $L$ as the independent variable and $T$ as the dependent variable and took into account the error on the temperature in the fitting.  The best-fit $L\propto T^4$ line results in a $\chi^2/{\rm dof}=28/5$ (probability $=3.6\times10^{-5}$) for the $L-kT$ diagram and a $\chi^2/{\rm dof}=41/5$ (probability $=9.4\times10^{-8}$) for the $L_{\rm disk}-kT$ diagram, respectively.  The $L\propto T^4$ relation is strongly excluded by the data in both cases. We note that if we use the $1\sigma$ error derived from the {\tt error} command in XSPEC, in which the $N_{\rm H}$ influence is not considered, the the best-fit $L\propto T^4$ relation will result in a $\chi^2/{\rm dof}=68/5$ (probability $=3\times10^{-13}$) for the $L-kT$ diagram.

We also performed a fit where the power-law index among $L$ and $T$ was left as a free parameter, $n$, with $L\propto T^n$.  The best-fit $L\propto T^n$ relations are shown as dashed lines in Fig.~\ref{fig:diskbb} with $n=-3.5\pm1.0$ for the $L-kT$ diagram and $n=-7\pm3$ for the $L_{\rm disk}-kT$ diagram.  These provide reasonable fits, with $\chi^2/{\rm dof}=2.7/4$ (probability = 0.61) for the $L-kT$ diagram and $\chi^2/{\rm dof}=5.9/4$ (probability = 0.21) for the $L_{\rm disk}-kT$ diagram. 

The best-fit disk inner radius $R_{\rm in}$ varies a lot between observations with different temperatures. Assuming a face-on disk at a distance of 4.13~Mpc, the best-fit $R_{\rm in}$ is about $(2-4) \times 10^4$~km for the low temperature group ($kT=0.11-0.12$~keV), but around $(1-2) \times 10^3$~km for the high temperature group ($kT=0.22-0.24$~keV).

\begin{figure}[t]
\centering
\plotone{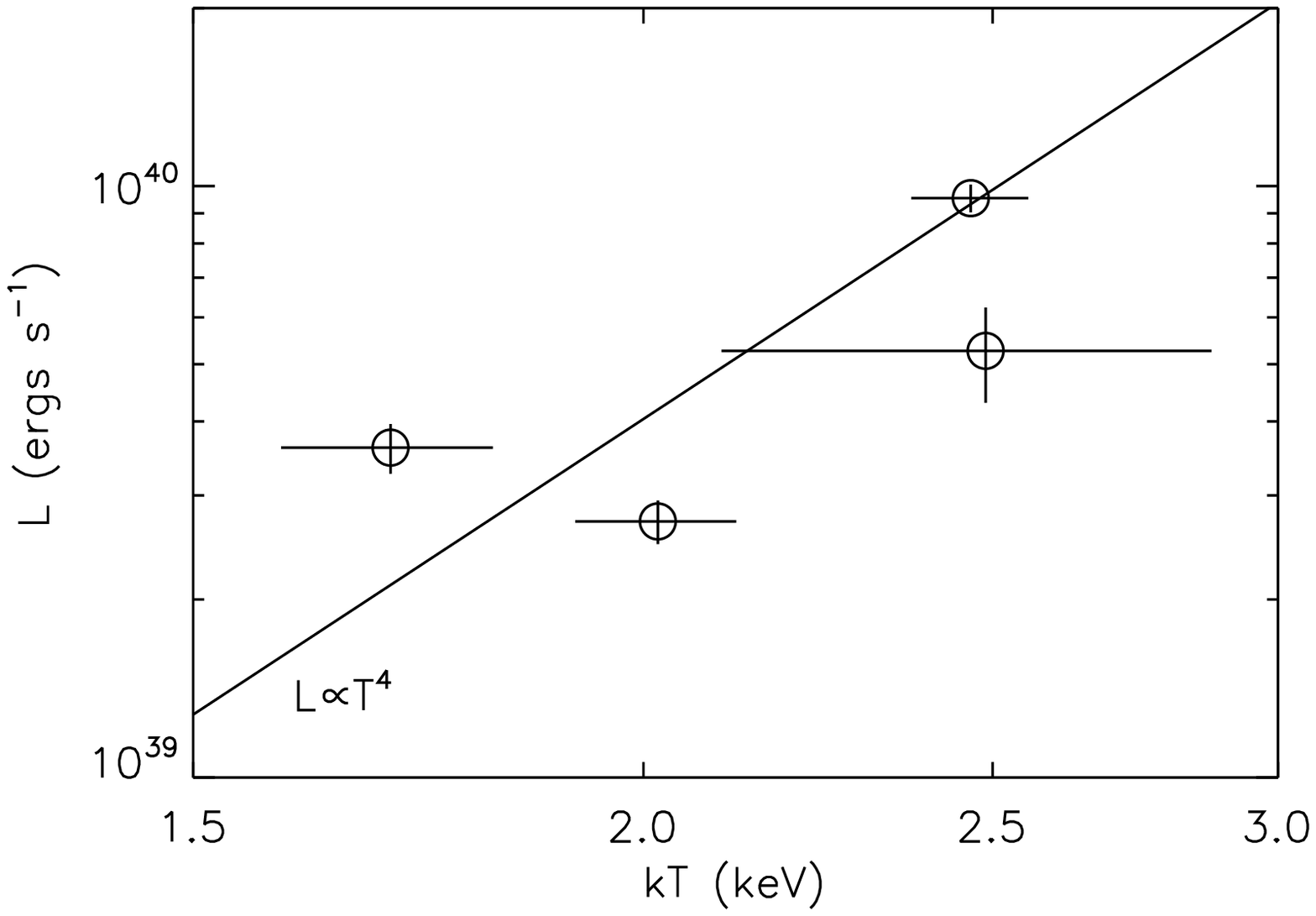}\\
\plotone{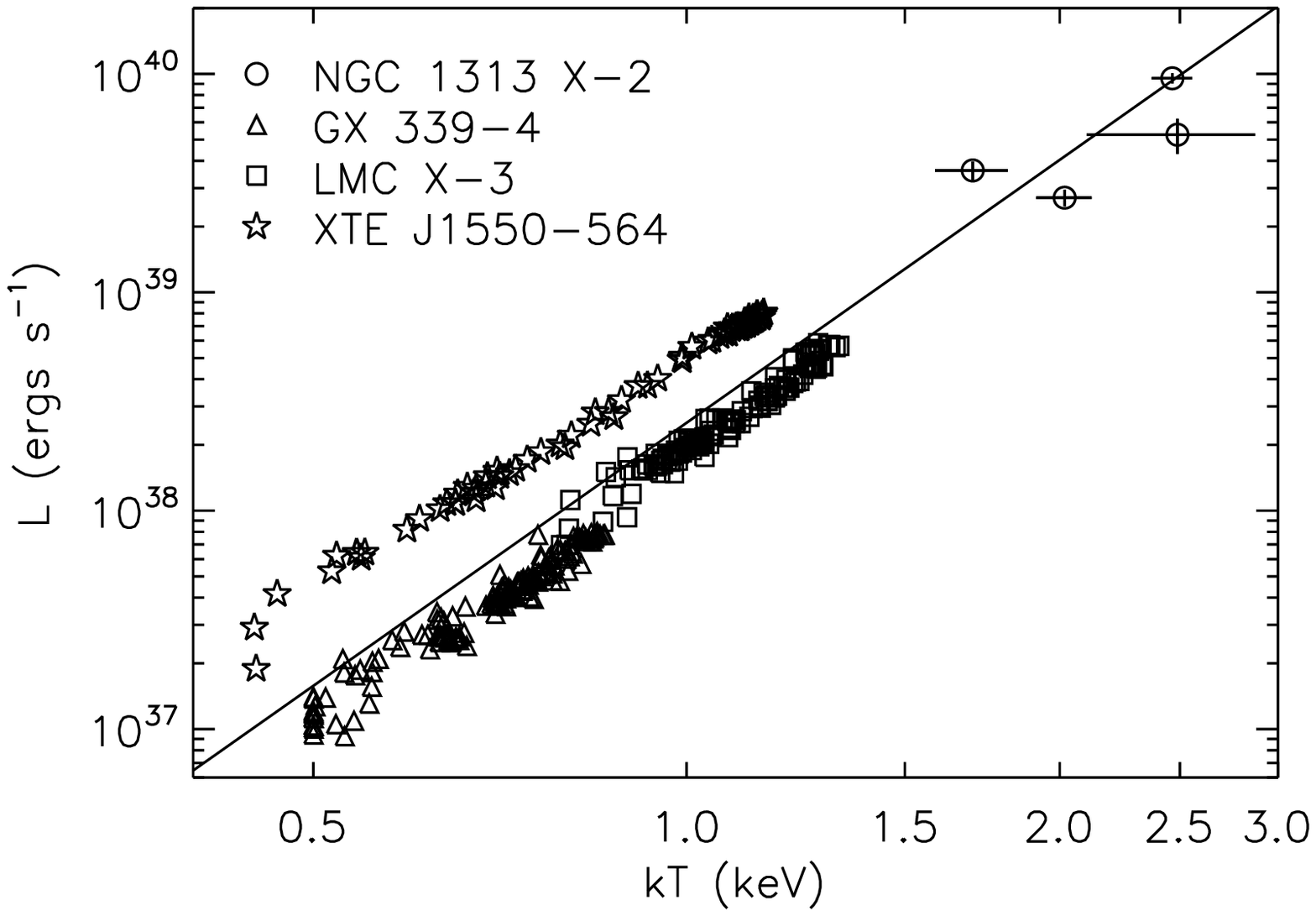}
\caption{
{\it Top}: $L-kT$ diagram of NGC 1313 X-2 for the $p$-free model from the four grouped {\it XMM-Newton} observations (for parameters, see Table~\ref{tab:spec}).  The solid line indicates the best-fit $L \propto T^4$ relation. 
{\it Bottom}: Same as the top panel but with $L-kT$ patterns of three Galactic black hole binaries: GX~339-4, LMC~X-3, and XTE~J1550$-$564 (data from \citealt{gie04a}).
\label{fig:pfree}}
\end{figure}

We fitted all spectra of X-2 with the $p$-free model. Due to the limited number of photons at high energies, errors on the temperature from a single observation are large, in particular for observations in the low luminosity state. We combined the 12 observations into 4 groups on basis of similar spectral properties, and performed simultaneous spectral fitting for each group to reduce the errors. Best-fit parameters for the 4 groups are listed in Table.~\ref{tab:spec}. We plotted the $L-kT$ diagram for $p$-free model in Figure~\ref{fig:pfree}.  The best-fit $L\propto T^4$ line is shown as a solid line in Figure~\ref{fig:pfree}. Although the 4 data points are insufficient to obtain an unambiguous pattern of the spectral evolution, they are roughly consistent with a $L\propto T^4$ relation, $\chi^2/{\rm dof}=9/3$.  In Figure~\ref{fig:pfree} (bottom), we overplotted the $L-kT$ patterns of three Galactic black hole binaries: GX~339$-$4, LMC~X-3, and XTE~J1550$-$564 (data from \citealt{gie04a}).

We also applied the MCD plus power-law model for the four grouped observations and found similar results with ungrouped data. Two of the four observations, group 1 and 2, present significant MCD components with temperatures of 0.22 and 0.11 keV, respectively, corresponding to a total 0.3--10 keV luminosity of $3.6\times10^{39}$ and $2.1\times10^{40}$ \ergs.  Again, the correlation between luminosity and temperature is opposite that expected if the soft thermal component arises from an accretion disk.

\begin{deluxetable*}{ccccccc}[ht]
\tablewidth{\textwidth}
\tablecaption{Best-fit spectral parameters of NGC 1313 X-2 with the $p$-free model
\label{tab:spec}}
\tablehead{
\colhead{group} & \colhead{observations} & 
\colhead{$N_{\rm H}$} & \colhead{$kT$} & \colhead{$p$} &
\colhead{0.3--10 keV $L$} & \colhead{$\chi^2/$dof}\\
\colhead{} & \colhead{} & 
\colhead{($10^{21}$ cm$^{-2}$)} & \colhead{(keV)} & \colhead{} & 
\colhead{($10^{39}$ ergs s$^{-1}$)} & \colhead{}\\
\colhead{(1)} & \colhead{(2)} & \colhead{(3)} & \colhead{(4)} & 
\colhead{(5)} & \colhead{(6)} & \colhead{(7)}
}
\startdata
1 & 1, 8, 10, 11 & 1.95$\pm$0.03 & 2.0$\pm$0.1 & 0.5000$^{+0.0003}_{-0.0000}$ & 2.7$\pm$0.2 & 986/838 \\
2 & 3, 4, 9, 12 & 2.18$\pm$0.07 & 2.5$\pm$0.1 & 0.589$^{+0.009}_{-0.004}$ & 9.5$\pm$0.5 & 2432/2420 \\
3 & 6, 7 & 2.22$\pm$0.05 & 1.7$\pm$0.1 & 0.500$^{+0.002}_{-0.000}$ & 3.6$\pm$0.3 & 477/461 \\
4 & 2, 5 & 2.39$\pm$0.09 & 2.5$\pm$0.4 & 0.501$^{+0.006}_{-0.001}$ & 5.2$\pm$1.0 & 560/409 \\
\enddata
\tablecomments{
Col. (1): Group index for combined spectral fitting.
Col. (2): Observation index quoted from \citet{fen06}.
Col. (3): Column density along the line of sight.
Col. (4): Disk inner temperature.
Col. (5): Power-law dependence of temperature on radius with a lower bounce of 0.5.
Col. (6): 0.3--10 keV intrinsic luminosity.
Col. (7): $\chi^2$ and degree of freedom.
All errors are at the 1 $\sigma$ level.}

\end{deluxetable*}

\section{Discussion}
\label{sec:diss}

The spectral evolution of NGC 1313 X-2 is inconsistent with the $L\propto T^4$ relation expected if the cool thermal component is, indeed, emission from an accretion disk at a high level of significance.  Detection of the cool disk component in ULXs is one of the arguments that support the presence of IMBHs. However, whether we can model the soft excess in the X-ray spectra of ULXs with a cool MCD, and even the reality of the soft excess itself, has been questioned.  \citet{gon06} found that in some bright ULXs the presence of the soft excess depended on how the power-law continuum was modeled. By selecting different energy ranges to fit the power-law continuum, the soft excess could be replaced by a soft deficit. Similar soft excesses have been found in many active galactic nuclei \citep[AGNs;][]{gie04b}, which can be phenomenologically modeled as blackbody emission with a temperature of 0.1--0.3 keV over a wide range of AGN luminosity and black hole mass. The disk emission in AGNs peaks in the ultraviolet band, unable to produce the soft excess peaked in the X-ray band. The soft excesses in ULXs are similar to those in AGNs, which suggests that they could have the same origin \citep{gon06}.  \citet{kin04} pointed out that mass outflow should occur in Eddington-limited accreting compact objects, and could produce the blackbody-like soft excess in ULXs and AGNs via a Compton thick envelope around the central object.  Therefore, it is imprudent to derive the mass of the compact object before we can determine the reality of the soft excess.

Spectral evolution offers a new means to test whether the soft excess is truly disk emission. The MCD dominates the high/soft spectral state of black hole binaries. Its evolution follows a $L\propto T^4$ pattern robustly and has been observed in many Galactic black hole binaries \citep{gie04a}.  In Figure~\ref{fig:diskbb}, either with the total luminosity or with the disk luminosity, the obvious departure from the $L\propto T^4$ relation with the MCD model places a strong constraint that the MCD model is not the correct interpretation of the soft excess in the spectra of NGC~1313 X-2.  The expected $L\propto T^4$ relation is very strongly inconsistent with the data.  Luminosities at high temperatures are more than one order of magnitude less than expected from the best-fit $L\propto T^4$. Although small deviations between $L$ and $T^4$ have been observed in Galactic black hole binaries, such large deviations have not been seen \citep{gie04a}.  The error on the temperature must be treated carefully because of the low flux of the MCD component and the fact that the MCD peak occurs where there is strong curvature in the spectrum due to interstellar absorption.  Our quoted uncertainties in the temperature take into account the possible effects of the absorption column density.  Even with these conservatively evaluated errors, the predicted $L-T^4$ relation is ruled out by the data at a confidence level of $4.1\sigma$.

The contraction of the disk inner radius at high temperatures is reminiscent of the ``anomalous regime'' in Galactic black holes GRO~J1655$-$40 \citep{kub01} and XTE~J1550$-$564 \citep{kub04}, where strong disk Comptonization takes place. However, the $L \propto T^4$ correlation for these sources is maintained if the total luminosity is considered instead of the disk luminosity. NGC 1313~X-2 fails to follow the $L \propto T^4$ for either the total or the disk luminosity.

\citet{fab01} have proposed that ULXs are similar to the Galactic Galactic X-ray binary SS~433, which has a powerful outflow with a kinetic power in the range $3 \times 10^{38}$ to $1 \times 10^{40}$ ergs s$^{-1}$, but viewed along the jet axis.  \citet{beg06} suggest that SS~433 is highly super-Eddington, by a factor of 5000, and that ULXs are also highly super-Eddington. \citet{pou07} considered the X-ray spectra in a similar model, also motivated by analogy with SS 433, and proposed that ULXs are geometrically beamed by a factor of 2--7, the soft excess is due to thermal emission from the spherization radius where the outflow originates, and the harder emission is from the accretion disk. Quasi-simultaneous observations made by {\it XMM-Newton} (part of these observations) and optical telescopes in 2004 showed that the optical counterpart of NGC~1313 X-2 did not brighten during the X-ray outburst \citep{muc06}, which might suggest that the outer region of the disk does not see the X-ray emission and that the X-rays are geometrically beamed.

In this scenario, the luminosity is not proportional to the temperature of the soft excess with a power of 4 as in the sub-critical accretion regime, instead, it varies inversely with the temperature of the soft excess, i.e., the temperature at the spherization radius $T_{\rm sp}$.  Our spectral results for NGC~1313 X-2 are consistent with the outflow model by \citet{pou07}, which show a correlation between $L$ and $T$ with a negative power-law index (Figure~\ref{fig:diskbb}).  Under this scenario, the soft excess is supposed to be modeled by a blackbody model instead of an MCD model.  We thus fitted spectra for the six observations with an absorbed blackbody plus power-law model, which also showed adequate fits. The $L-kT$ diagram for the blackbody model is shown in Figure~\ref{fig:bb}. The blackbody temperatures of around 0.1--0.2 keV indicate an accretion rate of about 20--100 times the Eddington rate and a spherization radius of about 200--1000 $GM/c^2$.  The best-fit dashed line indicates a power-law index of $-3.2$ between the two groups of data.  The trend between $L$ and $T$ is as expected for a super-Eddington accreting stellar-mass black hole, however, the slope is steeper than that predicted by \citet{pou07}.  

The spectral fitting with a $p$-free model is consistent with the $L\propto T^4$ relation. A recent {\it Suzaku} observation of NGC~1313 X-2 showing a $p$-free disk temperature of 1.86~keV with a luminosity of $10^{40}$ \ergs\ at a given distance of 4.13~Mpc \citep{miz07}, however, deviates from the best-fit $L\propto T^4$ line from {\it XMM-Newton} observations, which could be caused by the considerable error on the temperature or the difference between instrumental calibrations. In the $L-kT$ diagram (Fig.~\ref{fig:pfree}, bottom), NGC~1313 X-2 appears on the $L\propto T^4$ extension of the $L-kT$ patterns of Galactic black hole binaries. Again, this favors an interpretation in which the source is a stellar-mass black hole undergoing super-Eddington accretion.  Since the normalization of the $L\propto T^4$ line is dependent on the black hole spin, color correction factor, and temperature profile correction factor, the mass of the compact object cannot be precisely determined from Figure~\ref{fig:pfree}. We estimate the compact object mass of NGC~1313 X-2 to be around 10--20 $M_\sun$ as compared to Galactic black holes. 

We note that these results on NGC~1313 X-2 do not exclude the possibility that the soft excesses in other ULXs may be due to accretion disks.  As described in \citet{fen06}, the spectral evolution---even considering only the relation of the photon index versus luminosity of NGC~1313 X-2---is unlike that seen from stellar mass black holes.  Repeated observations of several ULXs will be required to measure the spectral variability patterns of a sample of ULXs to determine if the behavior of the soft excess is, in general, consistent with that expected for accretion disk emission.

\acknowledgments We thank Marek Gierli\'nski for providing the data of Galactic black hole binaries and the anonymous referee for the insightful comments, particularly regarding the uncertainty in the MCD temperature determination, which improved the paper significantly.

\end{document}